\documentclass[a4paper,fleqn]{cas-sc}

\usepackage[numbers]{natbib}
\usepackage{stfloats}

\newcommand*\dif{\mathop{}\!\mathrm{d}}

\begin{document}
\let\WriteBookmarks\relax
\def\floatpagepagefraction{1}
\def\textpagefraction{.001}

% Short title
\shorttitle{A Computational Analysis of Traffic Cluster Dynamics Using a Percolation-Based Approach in Urban Road Networks}    

% Short author
\shortauthors{Y. Kwon, M. Lee, M.J. Lee, and S.-W. Son}  

% Main title of the paper
\title [mode = title]{A Computational Analysis of Traffic Cluster Dynamics Using a Percolation-Based Approach in Urban Road Networks}  

% Title footnote mark
% eg: \tnotemark[1]
%\tnotemark[1] 

% Title footnote 1.
% eg: \tnotetext[1]{Title footnote text}
%\tnotetext[1]{} 

% First author
%
% Options: Use if required
% eg: \author[1,3]{Author Name}[type=editor,
%       style=chinese,
%       auid=000,
%       bioid=1,
%       prefix=Sir,
%       orcid=0000-0000-0000-0000,
%       facebook=<facebook id>,
%       twitter=<twitter id>,
%       linkedin=<linkedin id>,
%       gplus=<gplus id>]

\author[1]{Yongsung Kwon}%[<options>]

% Corresponding author indication
%\cormark[1]

% Footnote of the first author
%\fnmark[1]

% Email id of the first author
%\ead{}

% URL of the first author
%\ead[url]{}

% Credit authorship
% eg: \credit{Conceptualization of this study, Methodology, Software}
\credit{Conceptualization, Methodology, Software, Formal analysis, Investigation, Writing - Original Draft, Writing - Review \& Editing}
 
\author[2, 3]{Minjin Lee}%[<options>]

% Corresponding author indication
%\cormark[1]

% Footnote of the first author
%\fnmark[1]

% Email id of the first author
%\ead{}

% URL of the first author
%\ead[url]{}

% Credit authorship
% eg: \credit{Conceptualization of this study, Methodology, Software}
\credit{Conceptualization, Formal analysis, Investigation, Data Curation, Writing - Original Draft}

\author[3, 4]{Mi Jin Lee}[orcid=0000-0003-0013-7907]

% Corresponding author indication
\cormark[1]

% Footnote of the first author
%\fnmark[1]

% Email id of the first author
%\ead{}

% URL of the first author
%\ead[url]{}

% Credit authorship
% eg: \credit{Conceptualization of this study, Methodology, Software}
\credit{Conceptualization, Methodology, Formal analysis, Investigation, Data Curation, Visualization, Writing - Original Draft, Writing - Review \& Editing}

\author[1, 3, 5]{Seung-Woo Son}[orcid=0000-0003-2244-0376]

% Corresponding author indication
\cormark[2]

% Footnote of the first author
%\fnmark[1]

% Email id of the first author
%\ead{}

% URL of the first author
%\ead[url]{}

% Credit authorship
% eg: \credit{Conceptualization of this study, Methodology, Software}
\credit{Conceptualization, Methodology, Formal analysis, Investigation, Data Curation, Writing - Review \& Editing, Supervision}

% Address/affiliation
\affiliation[1]{organization={Department of Applied Artificial Intelligence, Hanyang University},
%            addressline={}, 
            city={Ansan},
%          citysep={}, % Uncomment if no comma needed between city and postcode
            postcode={15588}, 
%            state={},
            country={Korea}}
\affiliation[2]{organization={Research Center for Small Businesses Ecosystem, Inha University},
%            addressline={}, 
            city={Incheon},
%          citysep={}, % Uncomment if no comma needed between city and postcode
            postcode={22212}, 
%            state={},
            country={Korea}}
\affiliation[3]{organization={Department of Applied Physics, Hanyang University},
%            addressline={}, 
            city={Ansan},
%          citysep={}, % Uncomment if no comma needed between city and postcode
            postcode={15588}, 
%            state={},
            country={Korea}}
\affiliation[4]{organization={Department of Physics, Pusan National University},
%            addressline={}, 
            city={Busan},
%          citysep={}, % Uncomment if no comma needed between city and postcode
            postcode={46241}, 
%            state={},
            country={Korea}}    
\affiliation[5]{organization={Asia Pacific Center for Theoretical Physics},
%            addressline={}, 
            city={Pohang},
%          citysep={}, % Uncomment if no comma needed between city and postcode
            postcode={37673}, 
%            state={},
            country={Korea}}    
% Corresponding author text
\cortext[1]{mijinlee@pusan.ac.kr}
\cortext[2]{sonswoo@hanyang.ac.kr}

% Footnote text
%\fntext[1]{}

% For a title note without a number/mark
%\nonumnote{}

% Here goes the abstract
\begin{abstract}
Understanding the dynamics of traffic clusters is crucial for enhancing urban transportation systems, particularly in managing congestion and free-flow states. This study applies computational percolation theory to analyze the formation and growth of traffic clusters within urban road networks, using high-resolution taxi data from Chengdu, China. Presenting the road network as a time-dependent, weighted, directed graph, we identify distinct behaviors in traffic jam and free-flow clusters through the growth patterns of giant connected components (GCCs). A persistent gap between GCC size curves, especially during rush hours, highlights disparities driven by spatial traffic correlations. These are quantified through long-range weight-weight correlations, offering a novel computational metric for traffic dynamics. Our approach demonstrates the influence of network topology and temporal variations on cluster formation, providing a robust framework for modeling complex traffic systems. The findings have practical implications for traffic management, including dynamic signal optimization, infrastructure prioritization, and strategies to mitigate congestion. By integrating graph theory, percolation analysis, and traffic modeling, this study advances computational methods in urban traffic analysis and offers a foundation for optimizing large-scale transportation systems.
\end{abstract}

% Use if graphical abstract is present
%\begin{graphicalabstract}
%\includegraphics{}
%\end{graphicalabstract}

% Research highlights
%\begin{highlights}
%\item Simultaneous percolation analysis of traffic jams and free flow provides complementary insights into urban traffic dynamics.
%\item Gap metrics between traffic-jam and free-flow clusters reveal distinct growth patterns, particularly during rush hours.
%\item Correlation analysis highlights the influence of long-range interactions on traffic percolation thresholds and cluster formation.
%\item The road groups of weights show different correlations despite the same network.
%\end{highlights}

% Keywords
% Each keyword is seperated by \sep

\maketitle

% Main text
\section{Introduction}\label{sec:introduction}
Urban road systems are inherently complex, shaped by the interplay of countless vehicle movements, and influenced by a wide range of factors, including varying travel purposes~\cite{marta2012understanding, hamedmoghadam2021percolation}, the geographical and structural characteristics of road networks~\cite{kim2016exploring, sarkar2021road, wen2024optimizing}, and various external conditions~\cite{kwon2023global, serok2023enhancing}. This inherent complexity presents major challenges to maintaining smooth traffic flow and minimizing congestion in modern cities. Among the various outcomes of such complexity, traffic congestion stands out as a cascading process that significantly disrupts the functionality of road networks. A localized disturbance—such as an accident, sudden lane closure, or natural bottleneck—can propagate to adjacent areas, triggering a chain reaction of slowdowns. These effects are further intensified by spatial correlations between neighboring roads~\cite{helbing2001traffic, daqing2014spatial}, leading to the diffusion of local congestion into large-scale traffic jams characterized by long-range dependencies~\cite{wang2015percolation, petri2013entangled}.

Computational approaches have played a key role in deciphering the intricacies of traffic systems. Statistical physics, known for its capacity to model multi-particle interactions, offers a robust framework for understanding traffic behavior~\cite{chowdhury2000statistical, LEE20114555}. Complementarily, complex network analysis enables the examination of collective behaviors across transportation infrastructures, yielding insights into macroscopic traffic dynamics~\cite{BOCCALETTI2006175, ganin2017resilience, saberi2020simple, jung2023empirical}. These tools support the modeling of both localized disruptions and broader systemic behaviors within urban traffic systems. One particularly useful framework for analyzing such cascading phenomena is percolation theory. Originally developed in statistical physics~\cite{stauffer1994}, percolation theory has become a powerful tool for studying the robustness and vulnerability of complex networks~\cite{BOCCALETTI2006175, cohen2010complex}. In classical percolation models, nodes or links are randomly occupied, and a phase transition is observed when a giant connected component (GCC) emerges or fragments. The critical threshold marks the point at which a globally connected system becomes disintegrated. These principles, grounded in both graph theory and statistical mechanics, have been applied across various domains—including infrastructure resilience, epidemic spread, and interdependent systems.

More recently, percolation theory has been extended to urban traffic networks to model how free-flowing systems transition into large-scale congestion—and vice versa~\cite{li2015percolation, zhang2019scale, zeng2019switch}. Such analyses are effective in identifying connectivity thresholds and tracking the spread of congested regions~\cite{LI2015556}. These studies also revealed the structural dynamics of congestion~\cite{hamedmoghadam2021percolation, li2015percolation, zhang2019scale, zeng2019switch, ambuhl2023understanding, zeng2020multiple}, and highlighted the importance of long-range spatial correlations in the emergence of global traffic breakdowns~\cite{daqing2014spatial, zeng2020multiple, prakash1992structural, taillanter2021empirical}.
% However, applying classical percolation theory to empirical traffic data presents practical challenges: urban road networks are finite, inherently noisy, and characterized by complex spatial dependencies, which makes concepts such as universality and criticality are often less straightforward to apply. Consequently, recent studies have emphasized the identification of transition points as practical indicators of network fragility, rather than as strict manifestations of critical phenomena. 
Yet, most of those studies primarily focused on understanding the mechanism of the congestion process, either by analyzing the expansion of congested clusters~\cite{zhang2019scale} or the breakdown of free-flow clusters (i.e., the residual network after removing congested links)~\cite{zeng2019switch}. Various indices such as connectivity, robustness, and resilience, which have been developed to assess traffic network performance~\cite{LI20211} also only reflect the perspective of congestion formation. In contrast to the emphasis on how traffic systems collapse into congestion, the reverse process, how congestion dissipates and smooth traffic flow is restored, has received comparatively little attention. This aspect remains relatively underexplored, despite its central importance in maintaining stable, reliable, and resilient urban mobility systems.

To address this gap, our study simultaneously examines both free-flow and traffic-jam clusters within the same traffic configuration and provide a more integrated understanding of traffic dynamics shaped by congestion. Such dual observation is crucial for understanding urban road networks and thus enhancing the reliability of traffic evaluation systems because both phenomena occur concurrently and interact with each other. By leveraging high-resolution taxi data~\cite{guo2019urban} from Chengdu, China as a representative case, we apply percolation analysis to explore traffic dynamics to provide insights into the underlying mechanisms that can inform more resilient urban planning. Road segments are occupied sequentially based on their rescaled speed (high to low or vice versa), allowing for the simultaneous investigation of traffic-jam and free-flow processes. This approach provides a comprehensive view of urban traffic patterns, revealing disparities between peak and non-peak periods and offering new insights into the multifaceted features of traffic structure.

Our findings contribute to the development of more resilient traffic evaluation systems, minimizing congestion and improving urban road safety. By exploring the spatial correlations of traffic and their relationship with percolation patterns, we establish metrics that can guide dynamic traffic management and urban planning strategies. These insights are vital for designing robust transportation networks capable of maintaining functionality under stress.

The remainder of this paper is structured as follows: Section~\ref{sec:dataMethod} introduces the Chengdu road network and taxi dataset, along with the methods used for inputting traffic conditions and conducting traffic percolation analysis. Section~\ref{subsection:Percolation} presents an analysis of traffic-jam and free-flow percolation properties, while Section~\ref{subsection:Correlation} discusses the correlation patterns in the context of urban transportation. Finally, we summarize our findings and discuss their implications in the concluding section.

%Based on the analogy of \cite{zeng2019switch}, conventional percolation analysis of occupying congested roads can represent the transition process from the free state to the congested state. In contrast, the proposed free-flow percolation analysis assumes a scenario of congestion release, where a completely congested network gradually recovers, starting with the highest-speed roads within the traffic state. \noteMJL {is it correct?}\noteYK{In free-flow percolation in our case, better traffic condition road occupied first. It means that the original network is completely congested and we observe the good traffic cluster's growth. I think Ph.D. Minjin Lee creates nice ment. But actually I think the assumption that the completely congested (completely free-flow in traffic jam percolation case) is not essential. Because, we firstly focus on just growth GCC.} Note that our approach does not attempt to fully replicate the dynamic process of real-time congestion relaxation. 

\section{Computational Framework for Traffic Cluster Analysis Based On Percolation Process and Data Description}
\label{sec:dataMethod}

%\revise{\subsection{Percolation Theory In Urban Traffic}}
%\label{subsection:theory}
To explore these dynamic patterns, we now introduce the computational framework used in our analysis, including the structure of the dataset and our percolation-based methodology. Traditional percolation approaches based on random or sequential occupation can obscure meaningful dynamics, especially when the underlying weight distribution (e.g., traffic speed) is non-uniform. To address this, we adopt a modified percolation method that better reflects the structure and behavior of real traffic systems. In this Section, we describe the empirical dataset and detail the methodology used in our percolation analysis.

\subsection{Traffic-Jam and Free-Flow Percolation Analysis}
\label{subsection:method}

\begin{table*}[hb!]
\caption{Notation used in this paper. the notation ``Network'' describes the network structure, while ``Percolation'' refers to quantities obtained after applying the percolation process.}
\begin{tabular*}{\tblwidth}{@{}LL@{}}
%\textbf{Notation}             &                                                                                                \\
\toprule
\textbf{Network}              &                                                                                                \\
\midrule
$V$, $E$                           & a set of nodes (intersections) and a set of directed links (roads)                                                                 \\
%$E$                           & a set of directed links (roads)                                                                \\
$W(t)$                        & a set of link weights at time $t$                                                              \\
$G[V, E, W(t)]$               & the weighted directed traffic road network at time $t$                                                 \\
$N$, $L$                           & the number of nodes, $N=|V|$, and the number of directed links, $L=|E|$                                                                  \\
%$L$                           & the number of directed links, $L=|E|$                                                          \\
$r_i(t)$                      & the rescaled average speed of taxies on a road $e_i \in E$; i.e., $r_i(t)\in W(t)$                           \\
\midrule
\textbf{Percolation}          &                                                                                                \\
\midrule
$Q^{\rm(jam)}, Q^{\rm(free)}$ & the superscript notation; a quantity $Q$ in the traffic-jam (jam) or free-flow (free) cluster                                            \\
$p, p_c$                      & an occupation fraction of link and the transition fraction                                               \\
$s(p; t)$                     & the relative size $s$ of the giant connected component at occupation fraction $p$ and time $t$ \\
$\Delta A$                    & the gap area between two curves between $s^{\rm (jam)}(p)$ and $s^{\rm (free)}(p)$             \\
%$A$                           & the area under a curve                                                                         \\
$\alpha$                      & the exponent representing the long-rangeness of weight-weight correlation.\\
\bottomrule
\end{tabular*}
\label{tab:notation}
\end{table*}

\begin{figure}[hb]
\centering
\includegraphics[width=0.7\columnwidth ]{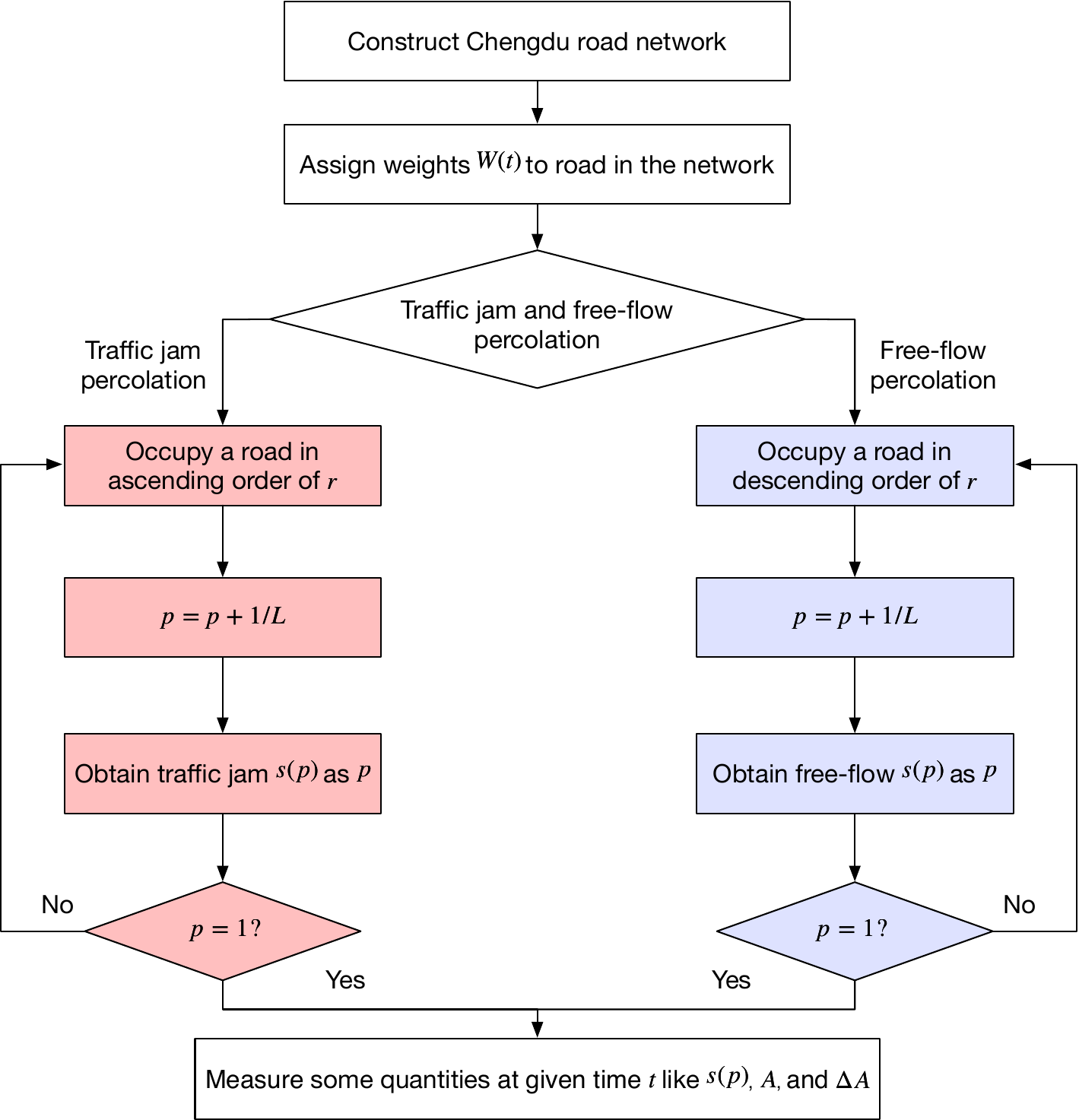} 
\caption{The flowchart describing the traffic percolation processes. The link density (occupation fraction) $p$ in the traffic-jam and free-flow percolation is explicitly $p^{\rm (jam)}$ and $p^{\rm (free)}$, respectively.}%
\label{Fig:flowchart}%
\end{figure}

%\subsubsection{Construction of the weighted directed network for traffic percolation analysis}
%\label{subsubsection:method_network}

To evaluate traffic conditions in urban road networks, we apply percolation theory to detect and analyze traffic clusters. The road network is modeled as a graph comprising  $N$ nodes (intersections) and $L$ directed links (roads) as an embedding structure. The notations used in this paper are summarized in Table~\ref{tab:notation}. 

As an indicator of the traffic status for a given road (link) $e_i \in E$, we use the average speed of the vehicle traveling along the road. However, raw speed data can vary significantly based on factors such as road type (e.g., highways versus alleyways) and daily traffic patterns, making direct comparisons impractical. To standardize these variations, we use the rescaled speed $r$, as proposed in previous research~\cite{li2015percolation}:
\begin{equation}
    r_{i}(t) \equiv \frac{v_{i}(t)}{v_{i}^\mathrm{max}},
\label{eq:rescaled_speed}
\end{equation}
where $v_{i}(t)$ is the average speed of vehicles on link $e_{i}$ at time $t$ and $v_{i}^\mathrm{max}=\max_t v_i(t)$ stand for the maximum observed speed on the same link during a given day. This standardized value allows us to express the relative traffic conditions of a given road within the range of 0 to 1 and serves as the link weight.

The resulting graph is a time-dependent, weighted, directed network $G[V, E, W(t)]$ (note that the rescaled speeds, namely the link weights set $W(t)$ can be varied in time $t$ while the embedding structure $V$ and $E$ remain static). This formulation allows for the dynamic modeling of traffic cluster formation and its temporal evolution, providing a robust computational framework for traffic assessment. We detect the traffic clusters using the percolation method. However, the random occupation used in the ordinary percolation is unsuitable for traffic analysis, where road usage is non-random and influenced by traffic flow dynamics. Thus, we adopt a weight-based percolation process tailored to traffic systems, following previous studies~\cite{li2015percolation, guo2019urban, RN95}, but introduce a modification to the role of the threshold value in an adaptive way.

In our approach, the weight $r$ of a link quantifies its traffic status, with values closer to 0 representing severe congestion (traffic jams) and values closer to 1 indicating smooth traffic flow (free flow). The occupation \emph{fraction} $p$ (not a probability as in ordinary percolation) as a control parameter determines the sequential occupation of links based on their weights. Links are occupied in ascending order (low to high $r$) for traffic-jam percolation and descending order (high to low $r$) for free-flow percolation. The sequential occupation of the lowest $r$ and highest $r$ corresponds to traffic-jam and free-flow percolation, respectively. Note that traffic clusters emerge in a deterministic manner due to the absence of stochasticity in the occupation process and that the two types of percolation are not exclusive, resulting in an overlap of the traffic-jam and free-flow components at large $p$. We observe cluster formation as a function of the occupied link density $p$ in each scenario. The entire procedure is illustrated in Fig.~\ref{Fig:flowchart}. With the percolation methodology defined, we next describe the urban environment and traffic data used to implement this framework.

\subsection{Data Description: Chengdu Road Network and Taxi Traffic}
\label{subsection:data}

\begin{figure}[b]
\centering
\includegraphics[width=0.8\columnwidth ]{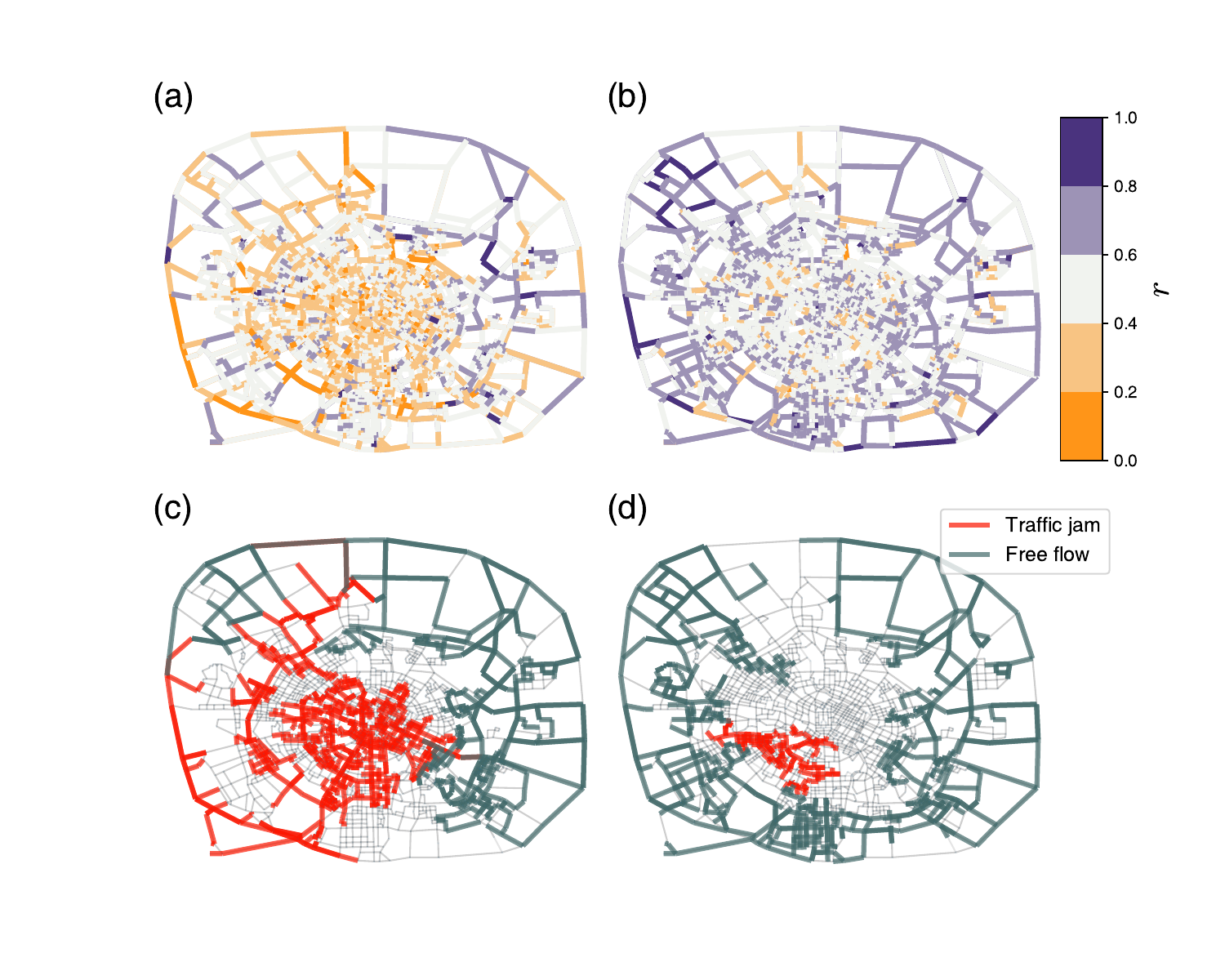} 
\caption{Traffic status of Chengdu road network. (a, b) Chengdu road networks denoted link weight as $r$ at (a) 08:00  and (b) 22:00 (Mon, June 15th, 2015). Roads that are closer to the orange indicate poor road conditions, while roads closer to the purple represent good road conditions. (c, d) The GCC of traffic jam (red edges) and free flow (green edges) when $p^{\rm(jam)}=0.3$ and $p^{\rm(free)}=0.3$, respectively, at (c) 8:00 and (d) 22:00 on the same day. The GCC of the traffic jam is formed near the center of the city, whereas that of the free flows occurs near highways outside the city.}%
\label{Fig:road_network}%
\end{figure}

\begin{table*}[hb!]
\caption{Road network and taxi speed data were collected in Chengdu, China~\cite{guo2019urban}. The taxi speed data can be accessed for various time windows as provided by the empirical data. The embedded road structure remains static.}
\begin{tabular*}{\tblwidth}{@{}LLLLL@{}}
\toprule
Period  & Weekday or Weekend & Time Windows (hour) & $N$ & $L$    \\
\midrule
June 1, 2015 to July 15, 2015 & \begin{tabular}[c]{@{}l@{}}32 weekdays\\ 13 weekend days\\ (total 45 days)\end{tabular} & \begin{tabular}[c]{@{}l@{}}03, 04, 08, 09, 12,\\ 13, 17, 18, 21, 22\end{tabular} & 1902 & 5943\\
\bottomrule
\end{tabular*}
\label{tab:data}
\end{table*}

This study utilizes the road network and taxi traffic data of Chengdu, China, as a representative case study~\cite{guo2019urban}. Chengdu, a metropolitan city with a population of approximately 16.33 million, provides a complex and diverse urban environment ideal for traffic-related research. The city’s road network is characterized by a central downtown core surrounded by a ring-shaped highway system. The Chengdu road network consists of $N=1902$ nodes (intersections) and $L=5943$ directed links (roads), forming the basis of our analysis. The road network provided by Ref.~\cite{guo2019urban} has been preprocessed and obtained from {\fontfamily{cmtt}\selectfont OpenStreetMap}~\cite{karduni2016protocol}.

To evaluate the traffic conditions of each road, we use processed taxi data from approximately 12,000 corporate taxis. This dataset includes detailed information on the time and speed of taxi movements on each road segment, and it covers a 45-day period from June 1 to July 15 in the year 2015. To mitigate noise caused by traffic signals or minor traffic events, we average taxi speed $v$ over one-hour intervals for each road segment. The preprocessing step ensures the reliability of the data for subsequent computational analysis. The taxi speed was assigned to the road segment by the {\fontfamily{cmtt}\selectfont Large-Scale Joint Map Matching algorithm} based on the Global Positioning System~\cite{li2013large}. 

The Chengdu road network, along with the rescaled speed data, is depicted in Figs.~\ref{Fig:road_network}(a) and~\ref{Fig:road_network}(b). These spatial visualizations highlight the distribution of well-performing roads (represented in purple) and poorly-performing roads (in orange) during both rush hour (8:00) and non-rush hour (22:00) periods. The rescaled speed $r$, as described in Sec.~\ref{subsection:method},  provides a standardized measure of relative traffic conditions, enabling consistent comparisons across road types and time periods. 

The original taxi data includes partial time stamps as listed in Table~\ref{tab:data}. For each day, this yields ten time points of rescaled average speed $r$ for every road segment. These data provide a robust temporal resolution for analyzing traffic dynamics and cluster formations, forming the foundation for the percolation-based methods employed in this study.

\section{Analysis of Traffic Clusters}
\label{sec:anlaysis}
Equipped with both our methodological framework and empirical dataset, we now proceed to a detailed analysis of the percolation patterns observed in the traffic network, mainly using the GCC growth curves.

\subsection{Giant Connected Components in Traffic-Jam and Free-Flow Clusters}
\label{subsection:Percolation}

\begin{figure}[t]
\centering % 
\includegraphics[width=0.73\linewidth]{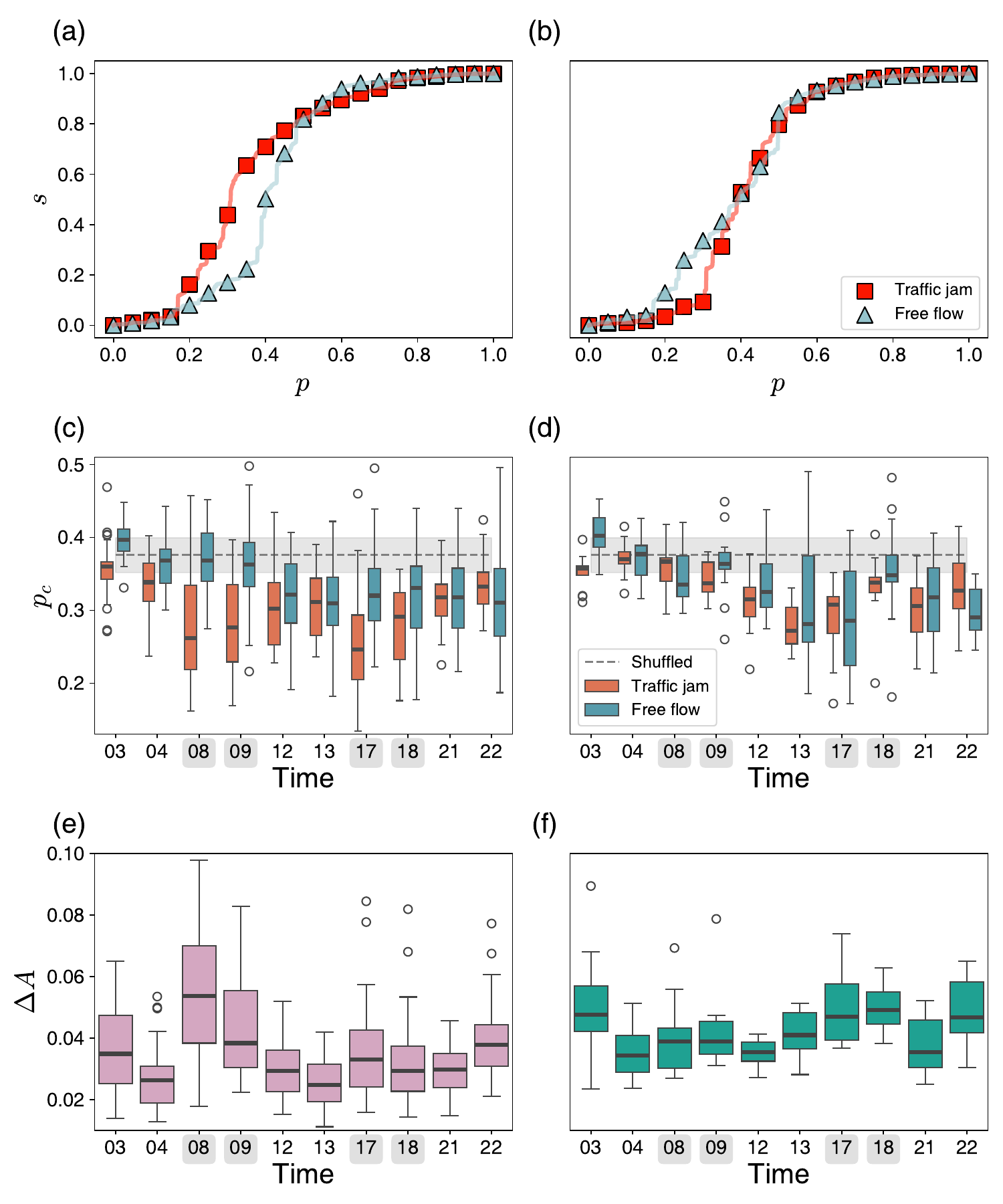}
\caption{The growth behavior of the GCC for traffic jam and free flow. Snapshots of the relative size $s$ of the GCC versus the occupation fraction (link density) $p$ (a) at 8:00 (morning rush hour) on Monday, June 15th, 2015 and (b) at 22:00 (non-rush hour) on the same day.  The $s^{\rm(jam)}$ and $s^{\rm(free)}$ at 22:00 behave similarly. The plot for the transition point $p_c$ versus time $t$ (c) for weekdays and (d) for weekends. The transition points $p_c^{\rm (jam)}$ and $p_c^{\rm (free)}$ are represented by the orange box and blue box, respectively, and the dashed line indicates the transition point on the weight-shuffled network, at $p_c^{\rm (shfl)}\simeq 0.376$. The plot for the gap area $\Delta A$ versus time $t$ (e) for weekdays and (f) weekends. The panels (c-f) display the quartile plot with the lower quartile $Q1$, the median or second quartile $Q2$, and the upper quartile $Q3$. The plot is a way to show the distribution of $p_c$ for (c), (e) 32 weekdays and (d), (f) 13 weekend days. The vertical length of the box stands for the interquartile range (IQR), $Q3-Q1$, the thick black line in a box indicates the median or the second quartile ($Q2$), and the whisker means the range between $Q1-1.5IQR$ and $Q3+1.5IQR$. Open circles out of the whisker correspond to the outliers. The gray box on the time axis is just for distinguishing two consecutive alternating time slots.}%
\label{Fig:GCC_gap_pc}
\end{figure}

In network theory, a giant connected component (GCC) represents the main structural unit of a system's functionality and robustness. In this study, we explore the formation and growth of GCCs under two distinct percolation scenarios: traffic-jam percolation, involving the sequential occupation of links in ascending order of weight $r$ (poor traffic conditions), and free-flow percolation, where links are occupied in descending order of $r$ (good traffic conditions). This process, illustrated in Fig.~\ref{Fig:flowchart}, forms the basis for understanding the dynamic interplay between traffic jams and free flow.

Spatial patterns of GCCs at the same link occupation density $p=0.3$ during  rush hour (08:00) and non-rush hour (22:00) are visualized in Figs.~\ref{Fig:road_network}(c) and~\ref{Fig:road_network}(d). More explicitly, the occupation density is denoted by $p^{\rm (jam)}$ for traffic jams and $p^{\rm (free)}$ for free flow. Despite the identical fraction $p$, the GCCs differ in spatial properties, such as size and shape, depending on the time of day and the percolation scenario. During rush hour, traffic-jam clusters are widespread across the urban center and extend to peripheral roads [Fig.~\ref{Fig:road_network}(c)], while free-flow clusters are usually confined to boundary regions [Fig.~\ref{Fig:road_network}(d)].   

Figures~\ref{Fig:GCC_gap_pc}(a) and~\ref{Fig:GCC_gap_pc}(b) depict the relative size $s(p;t)$ of the GCC as a function of $p$:
\begin{equation}
    s(p; t)\equiv \frac{S_1(p; t)}{N},
    \label{eq:gcc}
\end{equation}
for the representative cases on the same day as Fig.~\ref{Fig:road_network}. The $S_1(p; t)$ is the number of nodes (intersections) belonging to the GCC at occupation fraction $p$ at time $t$. During rush hour (8:00), $s^{\rm (jam)}$ grows earlier and more steeply than $s^{\rm (free)}$, reflecting the rapid propagation of congestion [Figs.~\ref{Fig:GCC_gap_pc}(a) and~\ref{Fig:GCC_gap_pc}(b)]. These behaviors suggest that traffic jams are more easily formed and spatially contagious compared to free flow, particularly during high-demand periods. Either the dominant traffic situation of traffic jams or free flow manifests in the form of an earlier onset and rapid growth of the GCC, in terms of percolation. All individual growth patterns of the GCCs for 45 days at a given time are plotted in Fig.~S1 in Supplemental Material.

\subsection{Transition Fraction and Correlation Analysis}
\label{subsubsec:pc}

While the previous section focused on the visual and growth patterns of traffic clusters, we now quantify these transitions through threshold analysis to better characterize system behavior. The transition fraction $p_c$, which signifies the onset of GCC formation, serves as a key indicator for analyzing traffic dynamics~\cite{zeng2019switch}. The smaller $p_c$ indicates the greater prevalence of the corresponding traffic clusters. In this empirical study, we evaluate the $p_c$ using both giant and second-largest components, following established percolation analysis techniques (see Fig.~S2 in Supplemental Material). Temporal variations in $p_c$ reveal remarkable insights into urban traffic dynamics, with clear differences between weekdays and weekends. The threshold values are compared against $p_c^{\rm(shfl)}\simeq 0.376$, obtained from weight-shuffled networks to account for the absence of spatial correlations. In Figs.~\ref{Fig:GCC_gap_pc}(c) and~\ref{Fig:GCC_gap_pc}(d), at first glance, one sees that the relations $p_c^\mathrm{(jam)}\lesssim p_c^{\rm(shfl)}$ and $p_c^\mathrm{(free)}\lesssim p_c^{\rm(shfl)}$ hold in most cases, which signals that the traffic correlation promotes the earlier onset of traffic clusters than expected. 

On weekdays, median transition values $\bar{p}_c$ (denoted by a thick gray line in the quartile plot) seem to behave in the opposite fashion for traffic-jam and free-flow percolation, with $\bar{p}_c^\mathrm{(jam)}<\bar{p}_c^\mathrm{(free)}$ kept. This pattern suggests that heavy-traffic roads cluster more easily into congestion components. During rush hours, the small $p_c^{\rm (jam)}$ highlights the rapid formation of traffic jams, consistent with previous studies in Beijing~\cite{li2015percolation,zeng2019switch}. Conversely, on weekends, similar trends in $\bar{p}_c$ for both scenarios suggest comparable ease in forming traffic-jam and free-flow clusters.

The temporal behavior of $\bar{p}_c$ further reflects daily traffic patterns. On weekdays, dips in $\bar{p}_c^{\rm (jam)}$ during morning and evening rush hours align with peak congestion [Fig.~\ref{Fig:GCC_gap_pc}(c)], as already revealed in the previous study~\cite{zeng2019switch}. In contrast, free-flow clusters show minimal sensitivity to rush hours, as indicated by relatively static  $\bar{p}_c^{\rm (free)}$ [Fig.~\ref{Fig:GCC_gap_pc}(c)]. Weekend activity patterns, concentrated around noon and late night, result in lower $\bar{p}_c$ at times such as 13:00 and 21:00 [Fig.~\ref{Fig:GCC_gap_pc}(d)].

\subsection{Gap Area and Growth Pattern Analysis}
\label{subsubsec:gap_area}
Beyond the threshold value alone, the structural discrepancy between traffic-jam and free-flow dynamics can be further understood by examining the area between their growth curves. The GCC growth pattern can be characterized by its point of onset and growth rate, and the small $p_c$ among these only ensures the early onset. The growth steepness can be evaluated by the critical exponent of the order parameter $s$ in percolation theory, but the exponent is neither analytically nor numerically ill-defined in these empirical data due to inherent fluctuation. To understand the entire behavior including the growth rate near the transition point $p_c$, we quantify the disparity between traffic-jam and free-flow clusters and calculate the gap area $\Delta A$ between the GCC growth curves, ranging from 0 to 1: 
\begin{equation}
    \Delta A \equiv \int_{0}^{1} \left|s^\mathrm{(jam)}(p) - s^\mathrm{(free)}(p)\right| \dif p.
\label{eq:gap_area}
\end{equation}
The gap area  $\Delta A$ captures the overall difference between the two processes, with $\Delta A=0$  representing identical curves and $\Delta A=1$ denoting the maximal dominance of either of the two curves. The latter arises when $s(p_c)=1$ at $p_c=1$ in one curve and $s(p_c)=1$ at $p_c=0$ in the other curve. 

Across all days and times, $\Delta A$ remains small (median around 0.03) but peaks during morning rush hours on weekdays ($\overline{\Delta A}\approx 0.05$) as seen in Figs.~\ref{Fig:GCC_gap_pc}(e) and ~\ref{Fig:GCC_gap_pc}(f). Taking into account the growth pattern of $s$ with increasing concave-down shape (Fig.~S1) as well as $\bar{p}_c^{\rm (jam)}<\bar{p}_c^{\rm (free)}$ at that time, we suspect that the quite long-lasting relation of $s^{\rm (jam)}>s^{\rm (free)}$ over all ranges of $p$ rather than $s^{\rm (jam)}<s^{\rm (free)}$ contributes to large $\Delta A$. 

\begin{figure}[b]
\centering % 
\includegraphics[width=0.8\columnwidth ]{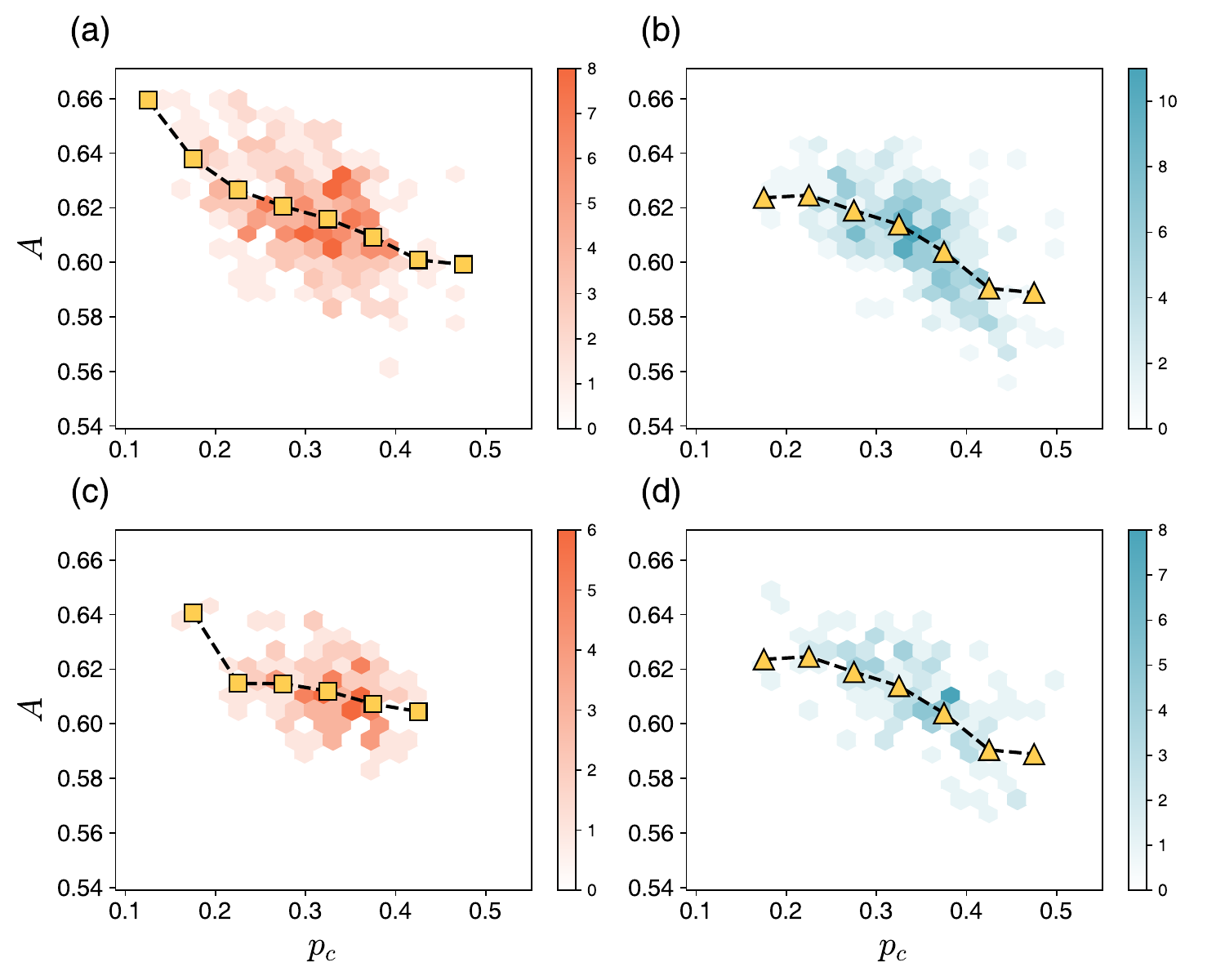}
\caption{Density plots of the area $A$ under the curve versus transition point $p_c$ on (a), (b) weekdays and (c), (d) weekends. The left panels (a), (c) and the right ones (b), (d) indicate the cases of traffic-jam percolation and free-flow percolation, respectively. The yellow squares and triangles represent the average values at a $p_c$. Overall, $A$ and $p_c$ are negatively correlated, with the Pearson correlation coefficients (a) $-0.49$, (b) $-0.63$, (c) $-0.36$, and (d) $-0.63$, respectively.}
%(a)week jam r=-0.4868, pv=1.890e-20 (b) week free r=-0.6289, pv=1.225e-36 (c) weekend jam, r=-0.3559, pv=3.244e-05 (d) weekend free, r=-0.6250, pv=1.901e-15}%
\label{Fig:pc_AUC}
\end{figure}

In most cases, $s$ increases trivially such as the rapid growth after $p_c$ and then the slow saturation at large $p$. Hence, early $p_c$ can guarantee fast arrival at $s\approx 1$ in this trivial growth pattern. The correlation between transition point $p_c$ and the area under the curve $A$, defined as \begin{equation}
    A \equiv \int_{0}^{1} {s(p) \dif p},
    \label{eq:auc}
\end{equation}
further characterizes GCC growth. Negative correlations between $p_c$ and $A$ are shown in Fig.~\ref{Fig:pc_AUC}, indicating that the early-onset of GCCs tend to grow rapidly and saturate, reflecting trivial growth patterns. However, nontrivial growth behaviors such as a long plateau at low $s$ are rare. This correlation also supports again the long-lasting relation $s^{\rm (jam)}>s^{\rm (free)}$ across threshold fraction $p$ at morning rush hours.

The combined use of $p_c$ and $\Delta A$ provides complementary insights into the simultaneous formation behaviors of traffic-jam and free-flow clusters. For instance, at evening rush hour (17:00) on weekdays, $\Delta A$ remains small despite a significant difference between $p_c^{\rm (jam)}$ and $p_c^{\rm (free)}$. This suggests that traffic-jam clusters during this period grow more slowly than in the morning, highlighting the complexity of urban traffic dynamics. Any single indicator alone, either of $p_c$ and $\Delta A$, is hard to perfectly explain the traffic situation, which is natural because the complex system is not simple. The effect of the difference itself as $p_c^{\rm (jam)}-p_c^{\rm (free)}$ is displayed in Fig.~S3 in Supplemental Materials.

\subsection{Correlation Analysis in Urban Traffic Patterns}
\label{subsection:Correlation}
%\subsubsection{Weight-weight correlation as a function of distance}
%\label{subsubsec:pcc}
\begin{figure}[]
\centering % 
\includegraphics[width=0.9\columnwidth]{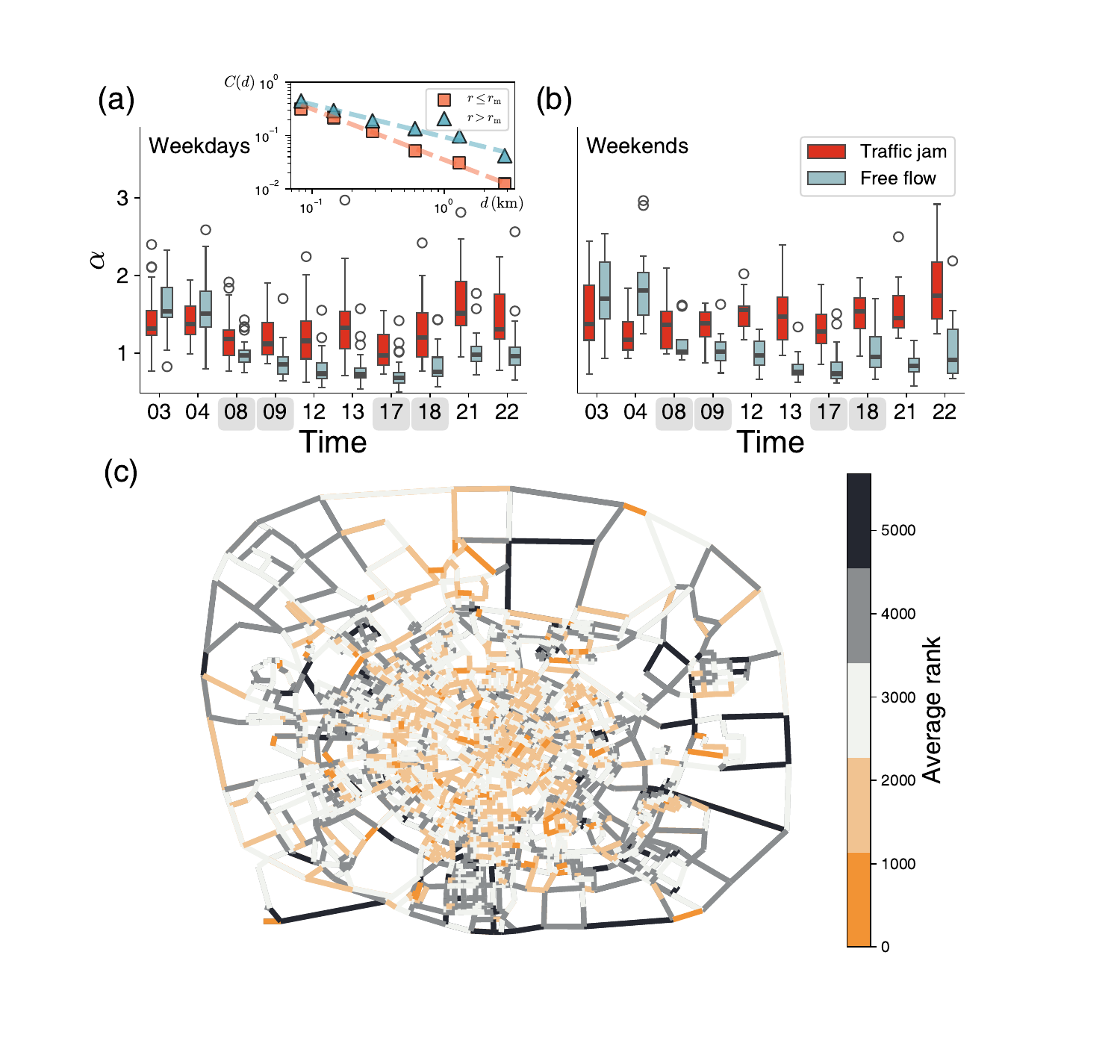}
\caption {The exponent $\alpha$ represents the long-rangeness of weight-weight correlation as time for (a) weekdays and (b) weekends. The orange quartile box plot represents the results where $r\le r_{\mathrm{m}}$ and the blue one represents the results where $r > r_{\mathrm{m}}$. (a) Inset: Weight-weight correlation as a function of the distance between two roads on June 4th, 2015, 17:00. The correlation of roads where $r\le r_{\mathrm{m}}$ is represented with an orange square, and the orange dashed line indicates the power-law fit. The blue triangle and dashed line are depicted as the correlation and power-law fitting line of roads where $r> r_{\mathrm{m}}$. An exponent of the power-law of $r\le r_{\mathrm{m}}$ is 0.96, and the one of $r>r_{\mathrm{m}}$ is 0.61. (c) Chengdu road network is depicted by the average rank of percolation occupation order for traffic-jam percolation. The roads that are closer to yellow are the ones occupied first in traffic-jam percolation, while the roads closer to black are the ones occupied first in free-flow percolation. In the case of traffic-jam percolation, the roads in the center of the city are preferentially occupied. In contrast, in the case of free-flow percolation, the roads on the outskirts of the city are occupied first. 
}
\label{Fig:alpha_time_map_rank}
\end{figure}

The traffic-jam and free-flow components emerge at different occupation fractions $p$ and grow differently. This variation may be influenced by underlying correlations in traffic conditions. Specifically, roads with similar traffic statuses---small $r$ for traffic jams and large $r$ for free flow---tend to spatially cluster more easily, facilitating the formation of the GCC. This hypothesis is supported by the observation that both $p_c^{\rm (jam)}$ and $p_c^{\rm (free)}$ are generally smaller than $p_c^{\rm (shfl)}$, as shown in Figs.~\ref{Fig:GCC_gap_pc}(c) and~\ref{Fig:GCC_gap_pc}(d). These findings align with previous studies demonstrating that positive (negative) spatial correlations promote earlier (later) GCC formation compared to uncorrelated structures in two-dimensional networks~\cite{prakash1992structural}. Motivated by this, we now turn to a quantitative examination of spatial correlation in traffic flows.

To measure this, we quantify the weight-weight correlation $C(d)$ between road segments as a function of geodesic distance $d$:
\begin{equation}
    C(d; t) = \frac{\sum_{ij} \big(r_{i}- \langle r \rangle \big) \big(r_{j} - \langle r \rangle \big) \delta(d_{ij} - d)}{\sqrt{\sum_{ij} \left[ \big(r_{i} - \langle r \rangle \big) \big(r_{j} - \langle r \rangle \big) \right]^2\delta(d_{ij} - d)}},
\label{eq:C_d}
\end{equation}
where $d_{ij}$ represents the geodesic distance (in integer bins) between roads $e_i$ and $e_j$, and $\langle r \rangle= \sum_i r_i(t) / L$. 

To characterize how long-range these correlations are, we assume a power-law form of the correlation as $C(d) \sim d^{-\alpha}$, where a smaller exponent $\alpha$ implies a stronger long-range correlation. This exponent serves as a proxy for what we coin the ``long-rangeness''. 

The previous findings~\cite{prakash1992structural} are summarized as follows: For $0 < \alpha < 2$, the correlation is positive, enhancing cluster formation, whereas $\alpha > 2$ corresponds to negative correlations. An exponent $\alpha = 2$ represents the absence of correlation. To estimate $\alpha$ for further clarity, we divide the data into two groups: roads with $r \leq r_m$ (traffic jams) and roads with $r > r_m$ (free flow), where $r_m$ is the median rescaled speed for a given day. As an example, the correlation $C(d)$ at 17:00 (evening rush hour) on June 4, 2015, follows a power-law decay with exponents less than 2, as shown in the inset of Fig.~\ref{Fig:alpha_time_map_rank}(a). Across most cases, $\alpha \lesssim 2$ for both weekdays and weekends, with correlations tending to weaken during the day and strengthen at night (see Supplemental Material, Fig.~S4).

Contrary to expectations, $\alpha$ is not lower for $r \leq r_m$ than for $r > r_m$ during rush hours, despite the earlier formation of congesting clusters (i.e., smaller $p_c^{\rm (jam)}$) in Figs.~\ref{Fig:alpha_time_map_rank}(a) and ~\ref{Fig:alpha_time_map_rank}(b). This can be explained by the spatial distribution of traffic-smoothing roads, which are typically located on the city’s outskirts and are longer in length, compared to congested roads concentrated in the urban center with shorter segments [Figs.~\ref{Fig:road_network}(c) and~\ref{Fig:road_network}(d)]. Traditional urban cities often exhibit a center-periphery structure~\cite{LEE2023113770, levinson2009minimum}, where the dense central network contrasts with the sparse, longer connections in the periphery. To confirm this, we computed the average rank of road occupation in ascending order over time [Fig.~\ref{Fig:alpha_time_map_rank}(c)]. Lower-ranked roads (congested) cluster in the center, while higher-ranked roads (smoothing) dominate the outskirts. These spatial characteristics suggest that $\alpha$ should be interpreted differently for traffic jams and free flow, particularly when considering road lengths.

%\subsubsection{Long-rangeness exponent and cluster formation pattern}
%\label{subsubsec:alpha}
\begin{figure}[]
\centering % 
\includegraphics[width=0.85\columnwidth]{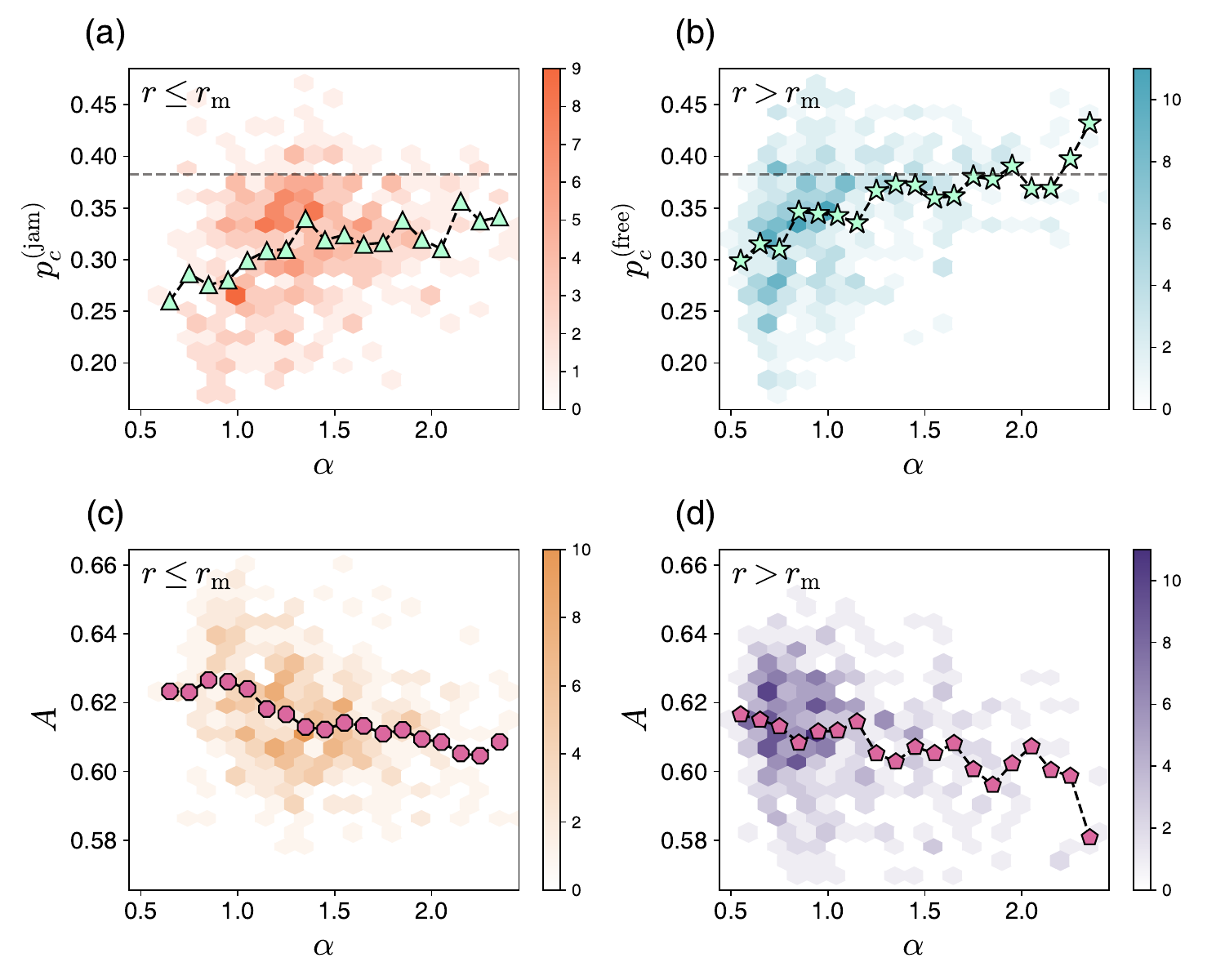}
\caption {The relation of the long-rangeness exponent $\alpha$ with (a), (b) the transition point $p_c$ and with (c), (d) the area $A$ under the curve, for the cases of (a), (c) $r\le r_{\mathrm{m}}$ and (b), (d) $r > r_{\mathrm{m}}$. The hexagonal tiles are the density plots, and other green and purple symbols with interpolation lines are average values at $\alpha$. In the upper panels (a) and (b), a dashed line represents $p_c^{\rm(*)}$ of random percolation on the Chengdu road network.}
\label{Fig:dvsC_aVSpc}
\end{figure}

We now explore how these correlation exponents relate to the percolation thresholds. Figures~\ref{Fig:dvsC_aVSpc}(a) and~\ref{Fig:dvsC_aVSpc}(b) reveal a positive relationship between $\alpha$ and the transition threshold point $p_c$, supporting the idea that stronger correlations reduce $p_c$ and facilitate the early prevalence of traffic clusters. The results align with the previous research~\cite{prakash1992structural} suggesting that long-range correlations (small $\alpha$) promote early GCC formation. These observations suggest that correlation plays a significant role in the early onset of traffic clusters, even though a rigorous theoretical framework may not be readily applicable. The higher $p_c^{\rm (free)}$ compared to $p_c^{\rm (jam)}$ at similar $\alpha$ values may be attributed to differences in network structures between the urban center and periphery. How can the relationship between the exponent $\alpha$ and the transition point $p_c$ be understood phenomenologically?

Consider traffic congestion: when one road becomes congested, adjacent roads are likely to follow, creating a cascading effect similar to a contact process. Speeds propagate across the network from the initial congested road. Stronger correlations (smaller $\alpha$) amplify this spread of similar traffic states, leading to the rapid formation of traffic clusters and a smaller $p_c$. This mechanism highlights how long-range correlations influence traffic dynamics by enabling the early and widespread formation of dominant traffic states, whether jams or free-flow conditions.

Long-range correlations not only drive early GCC onset but also accelerate their growth. To quantify this, we reuse the area $A$ under the growth curve $s(p)$, as defined in Eq.~(\ref{eq:auc}). Figures~\ref{Fig:dvsC_aVSpc}(c) and~\ref{Fig:dvsC_aVSpc}(d) show that $A$ decreases with increasing the longrangeness $\alpha$, reinforcing the hypothesis that long-range correlations (small $\alpha$) lead to rapid GCC growth. Median values of $A$ range from 0.56 to 0.67 for both $r \leq r_m$ and $r > r_m$, indicating substantial GCC development. Pearson correlation coefficients of -0.32 (traffic jams) and -0.22 (free flow) further confirm the negative relationship between $\alpha$ and $A$.

The observed relationships between small $p_c$, large $A$, and small $\alpha$ collectively highlight the influence of correlation on cluster dynamics. Long-range correlations enable the rapid spread of similar traffic states, such as congestion propagating from one road to its neighbors. This phenomenon is analogous to processes like contagion, where the prevalence of similar states spreads rapidly through a network. Together, these findings underscore the role of correlation in shaping dominant traffic conditions, whether as traffic jams or free-flow states.

\section{Conclusions}
\label{sec:conclusions}

As urban transportation systems grow increasingly complex, evaluating their dynamics requires a multidimensional approach rather than relying on a single metric. For instance, while some cities may recover quickly from significant congestion, others with generally smooth traffic flow might be highly vulnerable to minor disruptions. Unlike previous research that focused solely on traffic jams or free-flow phenomena, this study addresses this complexity by simultaneously examining traffic-jam and free-flow processes within urban road networks, providing a dual perspective on traffic dynamics. Using percolation analysis, we identify traffic clusters based on rescaled road speeds  (Fig.~\ref{Fig:flowchart}) and demonstrate the different behaviors of traffic jams and free flows during rush hours. The small $p_c$ values for traffic jams indicate the rapid formation of congestion, while the gap area $\Delta A$ provides additional insight into the differences between the two processes (Fig.~\ref{Fig:GCC_gap_pc}). Together, these metrics offer complementary tools for understanding urban traffic patterns more comprehensively. Additionally, our correlation analysis revealed that stronger weight-weight correlations can drive dominant traffic phenomena (Figs.~\ref{Fig:alpha_time_map_rank} and~\ref{Fig:dvsC_aVSpc}), emphasizing the role of network-wide interactions in shaping traffic behavior.

Our findings offer practical applications for improving the reliability and safety of urban road networks. Percolation-based methods can dynamically optimize traffic signal timings, mitigate congestion, and reduce accidents by leveraging real-time data. Emergency response systems could use these insights to identify and route through less congested and smoother paths, ensuring faster response times. Urban planners may prioritize infrastructure development in areas prone to congestion or design targeted interventions, such as congestion pricing or enhanced public transportation. Furthermore, real-time traffic management systems can redistribute vehicles more effectively, while predictive maintenance strategies can prevent infrastructure-related traffic disruptions. Smart parking solutions informed by traffic data can also minimize unnecessary driving, enhancing overall urban mobility. These practical applications underscore the value of our approach in supporting data-driven decision-making for urban transportation systems.

While this study focuses on Chengdu, the methodology is adaptable to other cities with different road network topologies. Chengdu’s core-peripheral ring structure differs from the lattice-like layouts of some planned cities, highlighting the importance of network topology in influencing traffic percolation and system resilience~\cite{LEE2023113770,LU2024110095}. Future work could extend this research by introducing theoretical modeling approaches and network-based traffic simulations. Correlated percolation models tailored to urban traffic conditions would allow for greater control over variables, enhancing our understanding of the interplay between network structure, correlation strength, and traffic dynamics. These advancements would provide a stronger foundation for developing predictive tools and more robust traffic management systems.

\section{Acknowledgments} \label{sec:acknowledgements}
This work was supported by the National Research Foundation (NRF) of Korea through Grant Numbers. NRF-2023R1A2C1007523 (S.-W.S.), RS-2024-00341317 (M.J.L.) and NRF-2022R1A5A7033499 (M.L.). This work was also partly supported by the Institute of Information \& communications Technology Planning \& Evaluation (IITP) grant funded by the Korean government (MSIT) (No.RS-2022-00155885, Artificial Intelligence Convergence Innovation Human Resources Development (Hanyang University ERICA)). We also acknowledge the hospitality at APCTP where part of this work was done.

% Numbered list
% Use the style of numbering in square brackets.
% If nothing is used, default style will be taken.
%\begin{enumerate}[a)]
%\item 
%\item 
%\item 
%\end{enumerate}  

% Unnumbered list
%\begin{itemize}
%\item 
%\item 
%\item 
%\end{itemize}  

% Description list
%\begin{description}
%\item[]
%\item[] 
%\item[] 
%\end{description}  

%\clearpage %%Remove this from your manuscript

% Uncomment and use as the case may be
%\begin{theorem} 
%\end{theorem}

% Uncomment and use as the case may be
%\begin{lemma} 
%\end{lemma}

%% The Appendices part is started with the command \appendix;
%% appendix sections are then done as normal sections
%% \appendix

%\section{}\label{}

% To print the credit authorship contribution details
\printcredits

%% Loading bibliography style file
%\bibliographystyle{model1-num-names}
%\bibliographystyle{cas-model2-names}
\bibliographystyle{unsrtnat}

% Loading bibliography database
%\bibliography{ref}

% Biography
%\bio{}
% Here goes the biography details.
%\endbio

%\bio{pic1}
% Here goes the biography details.
%\endbio

\section{Supplemental Material}
\appendix
\renewcommand{\thesection}{Note \arabic{section}}
\renewcommand{\thesubsection}{\arabic{subsection}}
\renewcommand{\figurename}{Figure}
\renewcommand{\tablename}{Supplementary Table} 
\renewcommand{\thefigure}{S\arabic{figure}}
\renewcommand{\thetable}{S\arabic{table}}
\graphicspath{{./}}
\setcounter{figure}{0}    
\setcounter{equation}{0}

%\pagebreak 

\section{Growth of traffic GCCs as the occupation fraction $p$}

\begin{figure*}[b]
\centering
\includegraphics[width=1\textwidth]{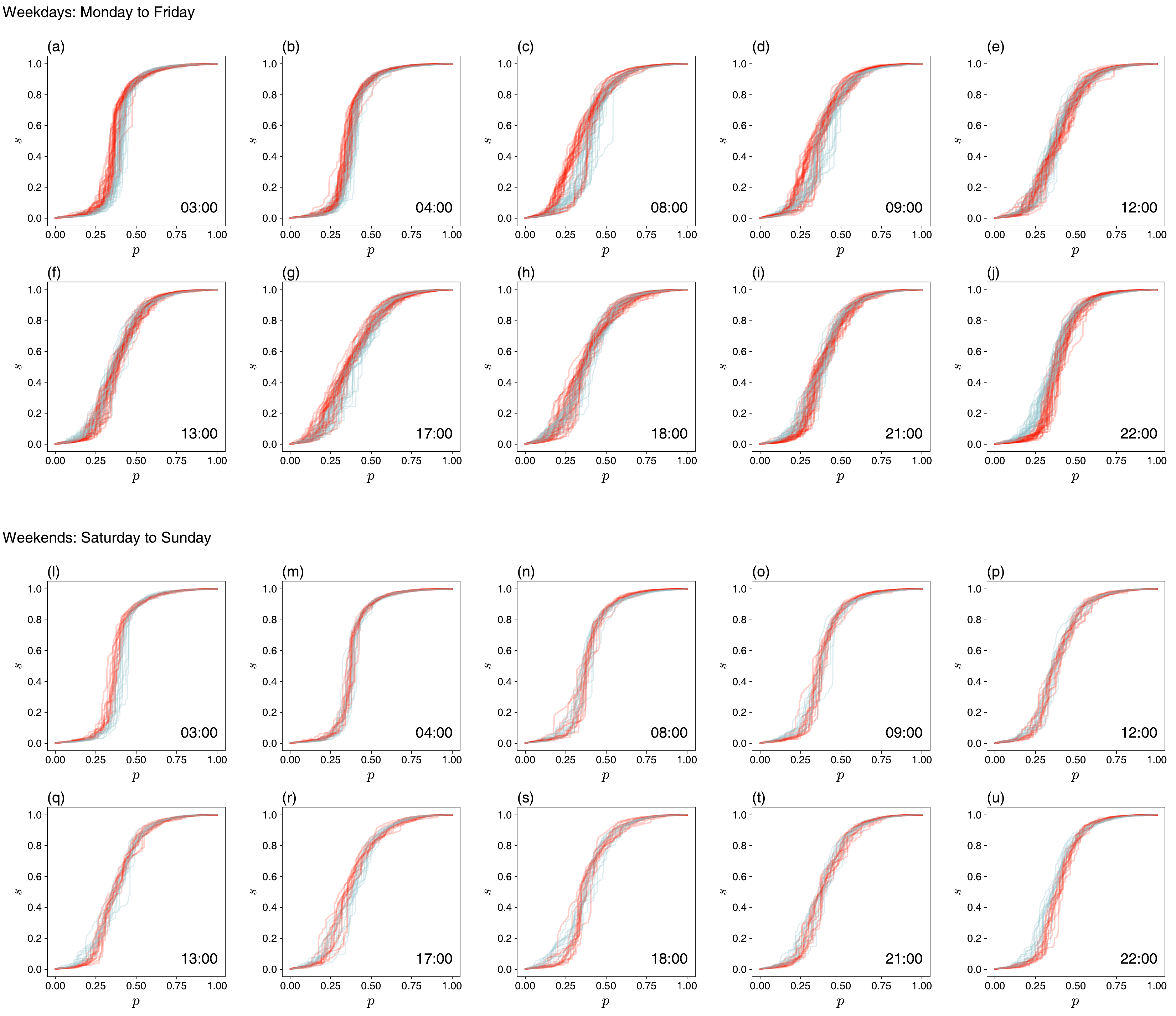}
\caption{The relative size of the GCC, $s_G$, versus occupation fraction $p$ for (a-j) 32 weekdays, (l-u) 13 weekends at a given time. Each panel corresponds to a given time slot, and the traffic-jam and free-flow GCCs are represented by red and blue lines, respectively.}
\label{suppfig:sp}
\end{figure*}

We obtain the relative size of the giant connected component curve as a function of the occupation fraction, which results in (two types of percolation)$\times(32 \mathrm{\, weekdays\,  or\, } 13 \mathrm{\, weekends})\times(\mathrm{ten\,  time-slots})$ curves $s_G(p)$ for our analysis. Every single curve is plotted in Fig.~\ref{suppfig:sp}. At a glance, almost all curves appear as increasing concave-down shapes. In shape, an early $p_c$ can ensure the early growth of $s$, resulting in a large area under the curve [Fig.~4 in the main text].

\section{Specifying percolation threshold $p_c$}
\begin{figure*}[b!]
\centering
\includegraphics[width=1\columnwidth]{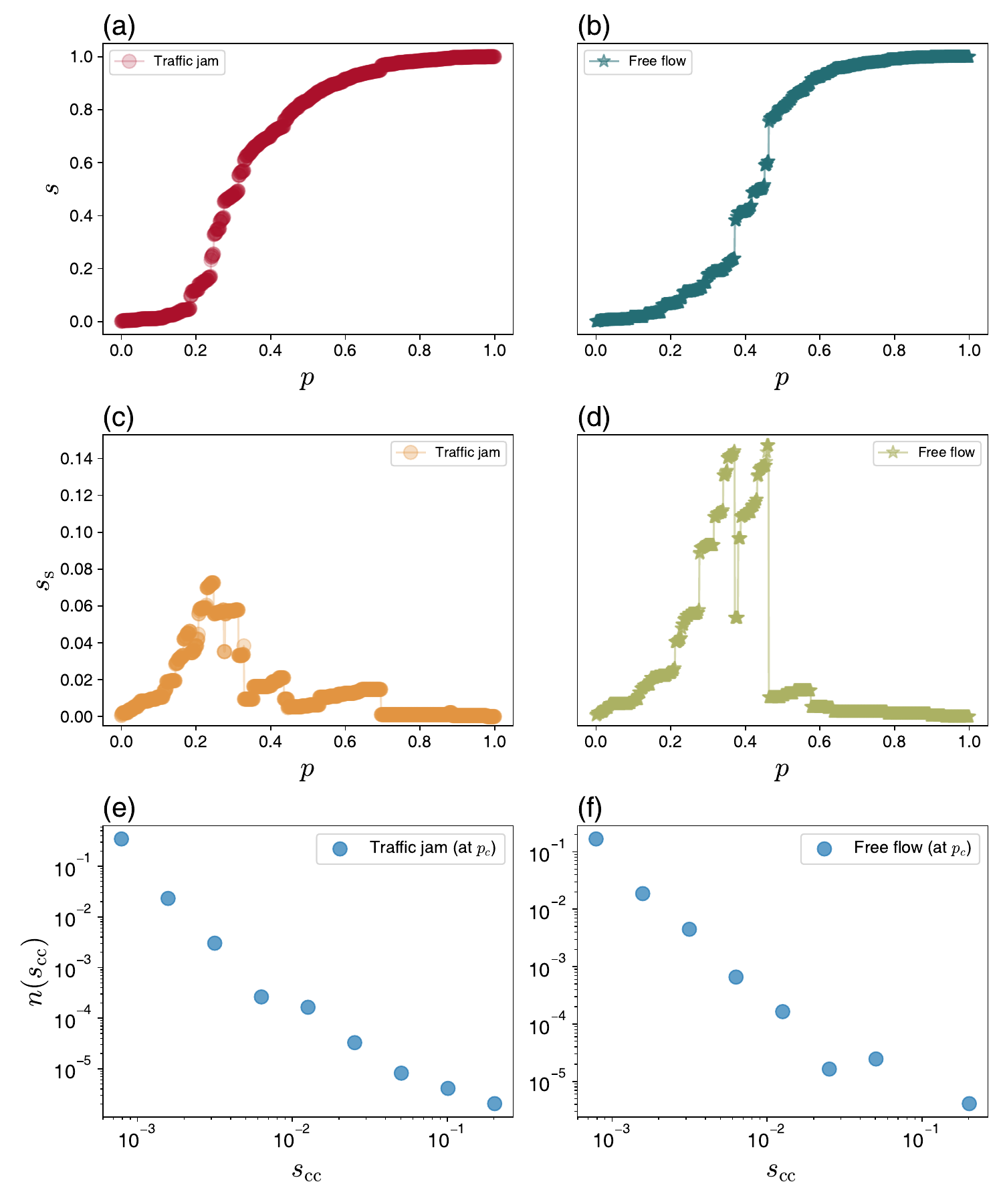}
\caption{Results of (a, c, e) traffic-jam and (b, d, f) free-flow percolations on June 4, 2015, 08:00. The relative size $s$ of the GCC of (a) traffic jam and (b) free flow behaves with increasing concave-down form. (c, d) The relative size $s_\mathrm{s}$ of the SGCC can have multiple peaks. (e, f) The cluster size distribution $n(s_\mathrm{cc})$ at $p_c$ follows a power-law form.}
\label{fig:GCC_SCC}
\end{figure*}

% 퍼콜레이션 이론에서 임계점은 가장 큰 클러스터의 크기가 시스템 크기에 준하는 크기로 창발하는 상전이 지점이다. 이 순간에 다양한 값들에서 scale-invariance한 현상들이 발견된다. 따라서 퍼콜레이션 이론에서 상전이 지점을 정의할 때, 다양한 값들의 scale-invariacne를 확인해야하지만, 실제 데이터에서는 데이터 수의 한계 때문에 정밀한 측정이 무의미할 수 있다. 따라서 실제 데이터에서 퍼콜레이션 분석을 다루는 몇몇 연구에서는 상전이 지점을 선정할 때 좀 더 쉬운 방법을 이용한다. 그중 하나가 GCC가 창발할 때 SCC가 GCC로 합병되는 지점이다. 이 순간과 이전 순간을 비교하면, SCC의 크기가 피크를 찍고 급격히 감소하는 경향을 보인다. 그러나 우리의 연구에서 다음 그림[1]과 같이 SCC의 픽이 여러개 찍히는 경우가 있다. 이를 GCC 크기와 비교하면 이 중 하나의 포인트를 선택하기는 애매한 경우가 있다. 맥시멈 지점을 고르더라고 GCC의 크기가 이미 시스템 크기에 준하는 크기가 되어 상전이 지점에 적합하지 않은 경우도 많다. 따라서 우리는 SCC가 이미 시템 크기에 준하는 GCC에 흡수되는 지점이 아닌, 시스템 크기에 준하지 못하는 크기의 커다란 클러스터들이 합쳐져 시스템 크기에 준하는 GCC가 발생하는 지점을 찾으려한다. 이러한 지점을 찾기 위해 SCC의 크기가 피크를 찍는 지점이며 GCC의 크기가 이미 커진 상태가 아닌 상황을 선정하기 위해 GCC<=0.5N인 조건을 적용했다.

% 그림[1]은 2015년 6월 4일 아침 8시의 데이터에서 퍼콜레이션을 진행한 결과이다. traffic jam percolation을 진행한 결과에서는 하나의 SCC 피크 지점을 고를 수 있지만, free-flow percolation을 진행한 결과에서는 SCC에서 2개의 피크 지점이 나타났다. 기존의 방법으로 critical point를 선택하면 SCC의 값이 더 큰 두번째 피크지점 (p=0.458)이 되겠지만, 이때 GCC의 크기는 0.8정도의 값이기 때문에 이미 시스템 크기에 준하고 있다. 따라서 우리는 값은 조금 작지만 GCC의 크기가 0.4 정도인 첫번째 피크지점 (p=0.370)을 critical point로 생각하였다. 

In percolation theory, the critical point $p_c$ is the phase transition point where the size of the GCC emerges, scaling with the size of the system. At this moment, various scale-invariant phenomena are observed, e.g., the divergence of the second giant connected component (SGCC) with a power-law form. This scale-invariant divergence cannot be obtained in a finite system, so the peak position (instead of the divergence point) of the SGCC has helped to measure $p_c$. However, in real-world data, multiple peaks of the SGCC exist due to their unavoidable innate fluctuations as well as finiteness, making it challenging to define only one $p$ that gives a significant single peak of SGCC as shown in Fig.~\ref{fig:GCC_SCC}. Let us denote $s$ and $s_s$ as the relative sizes of the GCC and SGCC, respectively. In the traffic-jam case, it can be said that there is only one large peak of $s_s$, but two peaks having similar $s_s$ seem to exist in the free-flow case. Which of the two peaks will give the threshold points?  

This observation drives us to define the threshold point in a more practical way confined to our traffic data. We impose the criteria in such a way that
\begin{equation}
    p_c \equiv \arg \max_p s_s(p) \text{  only\,if  } s(p)\leq 0.5.
    \label{eq:pc}
\end{equation}
The condition $s\leq 0.5$ is for ensuring that the GCC is not already significant when the $s_s$ peaks and resolves the vagueness like the abovementioned case in Fig.~\ref{fig:GCC_SCC}(d). Without this restriction, the pure maximum $s_s$ emerges at $p=0.458$. This $p$ gives the quite large value $s\approx 0.8$, and this large $s$ close to 1 does not describe the vicinity of the transition point. With the restriction $s\leq0.5$ imposed, the first peak of $s_s$ at $p=0.370$ gives $s\approx 0.4$, which sounds more reasonable. 

At the critical point, the cluster size distribution $n(s)$ follows the power-law form. We investigate the distribution $n(s)$ at the $p_c$ in Eq.~(\ref{eq:pc}). The distribution displayed in Figs.~\ref{fig:GCC_SCC}(e) and~\ref{fig:GCC_SCC}(f) follows a power-law form as expected.

\pagebreak

\begin{figure*}
\centering
\includegraphics[width=1\columnwidth ]{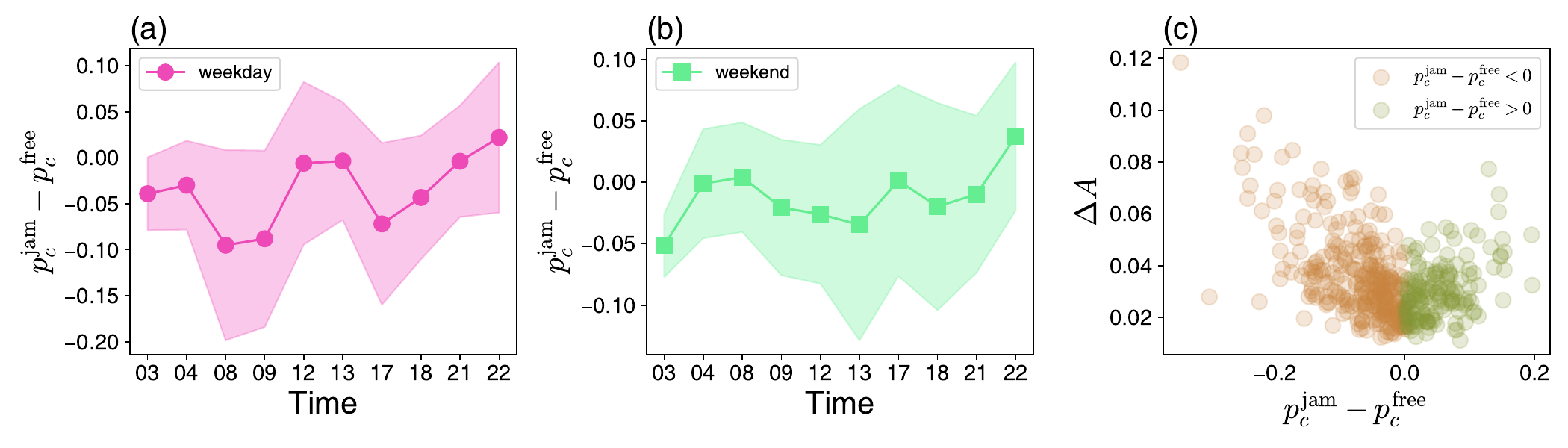} 
\caption{The behavior of the difference between $p_c^\mathrm{(jam)}$ and $p_c^\mathrm{(free)}$. (a, b)  The temporal change of the difference between $p_c^\mathrm{(jam)}$ and $p_c^\mathrm{(free)}$ on (a) weekdays and (b) weekends. The larger the magnitude of $p_c^\mathrm{(jam)}-p_c^\mathrm{(free)}$ is with a plus (minus) sign, the more likely the free-flow (traffic jam) GCC will occur. The difference can aid in figuring out traffic situations. (c) Then the scatter plot about $p_c^\mathrm{(jam)}-p_c^\mathrm{(free)}$ versus $\Delta A$. The Pearson correlation coefficient is estimated separately for $p_c^\mathrm{(jam)}-p_c^\mathrm{(free)}<0$ (brown) and $p_c^\mathrm{(jam)}-p_c^\mathrm{(free)}>0$ (green), as -0.61 and 0.44, respectively. Whether congestion or free flow is dominant, the gap area $\Delta A$ increases significantly as the degree of dominance grows. %Whether the congestion or free flow is dominant, the gap area $\Delta A$ increases significantly as the extent of the dominance grows.
}
\label{fig:pjam_pfree}%
\end{figure*}

\pagebreak

%\section{The exponent $\alpha$ of the long-range correlation}
\begin{figure*}
\centering
\includegraphics[width=1\columnwidth ]{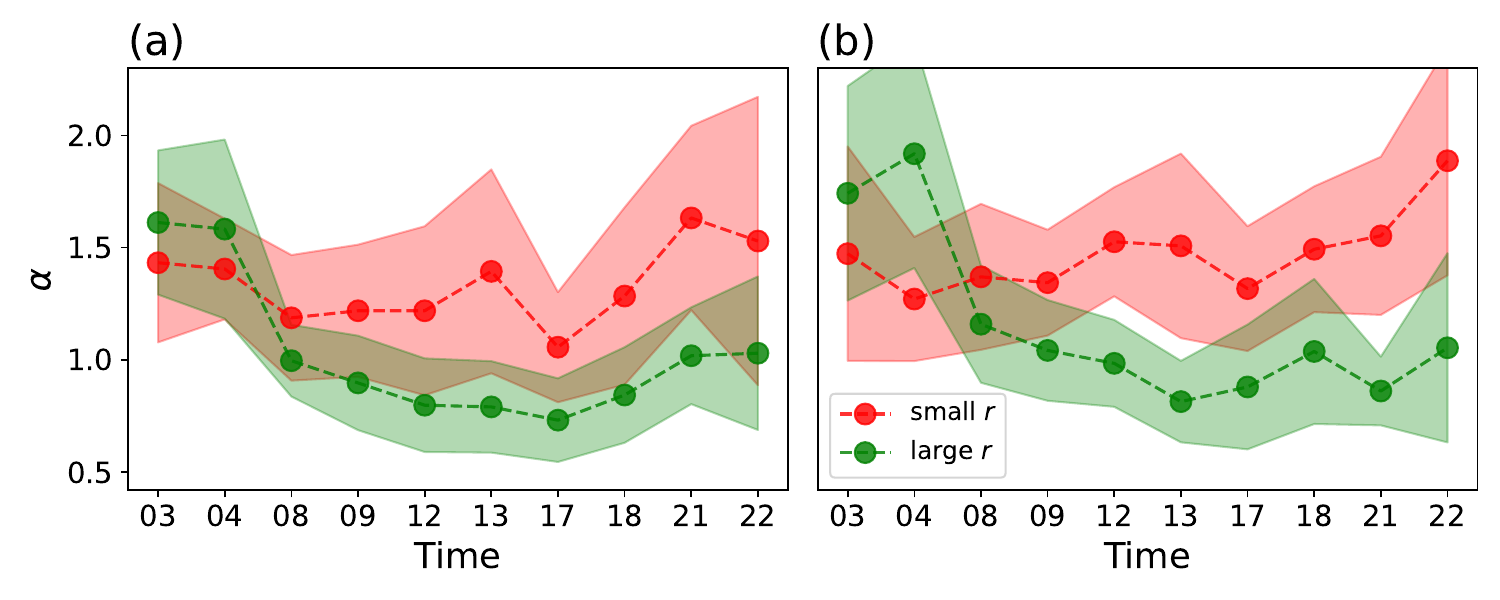} 
\caption{Temporal behavior of the long-rangeness exponent $\alpha$ on (a) weekdays and (b) weekends. The small $r$ means $r\leq r_\mathrm{m}$ while the large $r$ means $r>r_\mathrm{m}$ with $r_m$ being the median. $\alpha$ for the large-$r$ case is lower than that for the small-$r$ case, on average. $\alpha$ for the small-$r$ case seems to decrease during rush hours on weekdays. The exponent for the large-$r$ case decreases until the evening rush hour. On weekends, the $\alpha$ of small $r$ increases after evening rush hour. The $\alpha$ of large $r$ seems to behave like the case of the weekdays.}
\label{fig:time_alpha}%
\end{figure*}
%Figure~\ref{fig:time_alpha} shows the $\alpha$ as whole time. In the morning when people start their activity, both $\alpha_{r<r_\mathrm{median}}$ and $\alpha_{r>r_\mathrm{median}}$ decrease, but the $\alpha_{r>r_\mathrm{median}}$ shows smaller value than $\alpha_{r<r_\mathrm{median}}$. It means that good traffic conditions show a stronger correlation than bad traffic conditions. In commuting time (08:00 and 17:00 on weekdays), $\alpha_{r<r_\mathrm{median}}$ decreases, on the other hand, $\alpha_{r>r_\mathrm{median}}$ remains at the small value. Therefore, we can infer that rush hour congestion influences roads with bad traffic conditions.

\pagebreak

\end{document}